\documentclass[11pt,a4paper]{article}

\usepackage[utf8]{inputenc} 
\usepackage[T1]{fontenc}    %
\usepackage[english]{babel} 
\usepackage[final]{microtype} 

\usepackage{amssymb,amsmath,mathtools} 
\usepackage{hyperref}
\usepackage{amsthm} 

\usepackage{xcolor}
\usepackage{bbm}
\usepackage{braket}
\usepackage{url}

\usepackage[hmargin=0.12\paperwidth,vmargin=0.16\paperwidth,bindingoffset=0cm]%
{geometry} 

\usepackage{mathrsfs}

\theoremstyle{plain}

\theoremstyle{definition}

  \numberwithin{prop}{section}
   \numberwithin{cor}{section}
   \numberwithin{remark}{section}

\title{\Large\bfseries A Note on Optimal Soft Edge Expansions for the Gaussian $\beta$ Ensembles}%
   
\author{Peter J. Forrester${}^1$,
Anas A. Rahman${}^2$, and 
Bo-Jian Shen${}^1$}
\date{}

\begin{document}

\maketitle

${}^1$School of Mathematics and Statistics,  The University of Melbourne,
Victoria 3010, Australia. \: \: Email: {\tt pjforr@unimelb.edu.au};  {\tt bojian.shen@unimelb.edu.au}\\

${}^2$
Department of Mathematics,
The University of Hong Kong, Hong Kong. \\
Email: {\tt aarahman@hku.hk}

\bigskip

\begin{abstract}
\noindent 
We present some review material relating to the topic of optimal asymptotic
expansions of correlation functions and associated observables for
$\beta$ ensembles in random matrix theory. We also give an introduction to a related line of study that we are presently undertaking.
\end{abstract}

\vspace{.3cm}
\section{Global and Soft Edge Density for the GUE}\label{S1}
Consider first the Gaussian unitary ensemble (GUE) of $N \times N$ complex Hermitian matrices
$H = {1 \over 2} (G + G^\dagger)$, where $G$ is an $N \times N$
matrix of i.i.d.~standard complex Gaussian entries (also known as a complex Ginibre matrix, or a member of
the GinUE \cite{BF25}). Let $\rho_{(1),N}^{\rm GUE}(x)$ denote the eigenvalue density for GUE matrices, defined by the requirement that after integrating over  $(a,b)$, $a<b$, it is equal to the expected number of eigenvalues in this interval. It has been known for a long time \cite[\S 6.1.1]{Me67} that this eigenvalue density has the explicit functional form
\begin{align}
\rho_{(1),N}^{\rm GUE}(x) &= e^{- x^2} \sum_{j=0}^{N-1} {1 \over \sqrt{\pi} 2^j j!} (H_j(x))^2 \nonumber
\\&= {e^{-x^2} \over \sqrt{\pi} 2^N (N-1)!}
\Big ( H_{N-1}(x) H_N'(x) - H_{N-1}'(x) H_N(x) \Big ),\label{0.1}
\end{align}
where $H_j(x)$ denotes the Hermite polynomial and the second equality follows from a confluent form of the Christoffel-Darboux formula; see, e.g., \cite[Eq.~(5.13)]{Fo10}. Plots of the difference (\ref{0.1}), which depends only on the $(N-1)$- and $N$-th Hermite polynomials, for increasing $N$ show that, to leading order, the density has a semi-circle profile supported on $(-\sqrt{2N}, \sqrt{2N})$. Plotting the normalised scaled quantity 
$\bar{\rho}_{(1),N}^{\rm GUE,g}(X) := {\sqrt{2N} \over N} \rho_{(1),N}^{\rm GUE}(\sqrt{2N}X)$ shows
compelling evidence for the limit law
\begin{equation}\label{0.2}
\lim_{N \to \infty} \bar{\rho}_{(1),N}^{\rm GUE,g}(X) = \bar{\rho}_{(1),\infty}^{\rm W}(X),
\qquad \bar{\rho}_{(1),\infty}^{\rm W}(X) := {2 \over \pi} (1 - X^2)^{1/2} \mathbbm 1_{|X|<1},
\end{equation}
where $\bar{\rho}_{(1),\infty}^{\rm W}(X)$ is known as the Wigner semi-circle. In fact,
(\ref{0.2}) was established by Wigner without knowledge of the explicit functional form (\ref{0.1}) in a pioneering analysis based on the method of spectral moments
\cite{Wi58}.

By the scaling used to define $\bar{\rho}_{(1),N}^{\rm GUE}(X)$ --- referred to as the global scaling --- the eigenvalues are  to leading order supported on the compact interval $(-1,1)$, as is clear from (\ref{0.2}). Starting with \cite{Fo93a}, much attention has been focused on the neighbourhood of the largest eigenvalue, and a particular scaling --- referred to as the soft edge scaling --- which shifts the origin to the leading order location of the largest eigenvalue (this being $\sqrt{2N}$) and then introduces a scale so that the mean spacing between eigenvalues is unity by the mapping
$x \mapsto \sqrt{2N} + y/(\sqrt{2} N^{1/6})$ (here the factor $\sqrt{2}$ multiplying $N^{1/6}$ is just for convenience). Making use of (\ref{0.1}) and known asymptotic formulae for the Hermite polynomials, it was established that the eigenvalue density with this scaling obeys the limit law
\begin{align}
\lim_{N \to \infty} 
\rho_{(1),N}^{\rm GUE,s} (y)
&\;= - y
({\rm Ai}(y))^2 + ({\rm Ai}'(y))^2,\nonumber
\\\rho_{(1),N}^{\rm GUE,s} (y)
&:= {1 \over \sqrt{2} N^{1/6}}
\rho_{(1),N}^{\rm GUE} (\sqrt{2N} + y/(\sqrt{2} N^{1/6}) ),\label{0.3}
\end{align}
where ${\rm Ai}(y)$ denotes the Airy function.

Taking a viewpoint of the rate of convergence in probability theory, or that of asymptotic expansions in applied mathematics, one would like to specify not only the limit laws for $\bar{\rho}_{(1),N}^{\rm GUE}(X)$ and 
$\rho_{(1),N}^{\rm GUE,s} (y)$, but also correction terms as a function of $N$. In the case of the density with global scaling this question needs to be posed as asking for the large $N$ expansion of
the smoothed quantity
$\int_{-\infty}^\infty f(X) 
\bar{\rho}_{(1),N}^{\rm GUE,g}(X) \, dX$ for a suitable class of test functions $f(X)$, due to the next term in the pointwise correction to the Wigner semi-circle involving a factor of $\cos (2 N \pi P(X))$, 
$P(X) := 1 + (X/2)\bar{\rho}_{(1),\infty}^{\rm W}(X) - (1/\pi) {\rm Arccos}(X) $
\cite{KB02}. In keeping with Wigner's study of the limiting global density through its spectral moments, choosing for $f(X)$ the even monomials $f(X) = X^{2k}$, $(k=1,2,\dots)$, for large $N$ it is a celebrated result in random matrix theory that the large $N$ expansion is in powers of $1/N^2$ and terminates \cite{BIPZ78}:
\begin{equation}\label{0.4}
m_{2k}^{\rm GUE} := 
\int_{-\infty}^\infty X^{2k} 
\bar{\rho}_{(1),N}^{\rm GUE,g}(X) \, dX
= \sum_{j=0}^k {a_{2j}(2k) \over N^{2j}},
\end{equation}
for some coefficients $\{ a_{2j}(2k) \}$
with a combinatorial/ topological meaning
(for example, after multiplying by
$2^k$, $a_{2j}(2k) |_{j=0}$ is the $k$-th Catalan number, as already deduced by Wigner \cite{Wi58}). As with Wigner's work, the analysis leading to this made no use of (\ref{0.1}), rather the result was obtained by introducing a graphical calculus associated with the definition of the moments as matrix averages.

Turning to the situation with soft edge scaling, making use of (\ref{0.1}),
the work \cite{GFF05} gave the large $N$ expansion
\begin{multline}\label{0.5}
\rho_{(1),N}^{\rm GUE,s} (y) =
\rho_{(1),\infty}^{\rm GUE,s} (y) 
-{1 \over 20} \Big (
3 y^2 ({\rm Ai}(y))^2 - 2 y ({\rm Ai}'(y))^2 - 3 {\rm Ai}(y) {\rm Ai}'(y) \Big ) {1 \over N^{2/3}} 
\\+ {\rm O} (N^{-1}),
\end{multline}
where $\rho_{(1),\infty}^{\rm GUE,s} (y)$ is the limiting functional form given in the first equation of (\ref{0.3}). In the follow up work 
\cite[Eq.~(2.16)]{FFG06e}, an explicit functional form was stated for the ${\rm O} (N^{-1})$ term in (\ref{0.5}). 
Unlike (\ref{0.4}), which has been the subject of sustained interest since its discovery,
it is only in recent years that higher order terms in (\ref{0.5}) have again been the subject of interest in the literature \cite{Bo24,Bo25a}. From this one learns that the stated functional form at ${\rm O} (N^{-1})$ stated in
\cite[Eq.~(2.16)]{FFG06e} is actually in error and there is in fact no such term.
Rather, the next order term occurs at order $N^{-4/3}$, and equals
\begin{equation}\label{0.6}
 {1 \over 16} \bigg ( 
\Big ( {39 y^3 \over 175} + {9 \over 100 } \Big ) ({\rm Ai}(y))^2 - {3 y^2 \over 175} ({\rm Ai}'(y))^2 -
\Big ( { y^4 \over 25} + {99 y\over 175 } \Big ) {\rm Ai}(y) {\rm Ai}'(y) \bigg ) 
{1 \over N^{4/3}}.
\end{equation}
Here, a remarkable structure is visible, namely that this involves a linear combination of the transcendental basis
$\{ ({\rm Ai}(y))^2, ({\rm Ai}'(y))^2,
{\rm Ai}(y) {\rm Ai}'(y) \}$, in which the coefficients are (low order) polynomials. Furthermore, checked to the first 10 orders in \cite{Bo24} is that the GUE correlation kernel
\begin{equation*}
K_N^{\rm GUE}(x,y) := e^{-(x^2+y^2)/2}
\sum_{j=0}^{N-1} {1 \over \sqrt{\pi}2^jj!} H_j(x) H_j(y)
\end{equation*}
has with respect to both variables $x,y$ a large $N$, soft edge variables form, which at each order $N^{-2j/3}$, $j=1,2,\dots$,
consists of a linear combination of the basis functions 
\begin{equation*}
\{ {\rm Ai}(x) {\rm Ai}(y), {\rm Ai}'(x) {\rm Ai}'(y),
{\rm Ai}(x) {\rm Ai}'(y), {\rm Ai}'(x) {\rm Ai}(y) \},
\end{equation*}
with coefficients that are polynomials in both $x$ and $y$. Since $\rho_{(1),N}^{\rm GUE} (x) = K_N^{\rm GUE}(x,x) $, this establishes that higher order terms in
(\ref{0.5}) all occur at powers of $N^{-2/3}$, and with a structure involving the transcendental basis noted below (\ref{0.6}) with polynomial coefficients, at least to the first 10 orders. An aim of our research is to give a different viewpoint on this and related results that we now turn to.

\section{Global and Soft Edge Density for the GOE and GSE}
Here, we wish to highlight that the large $N$ expansions of the global and soft edge scaled densities for the Gaussian orthogonal ensemble of real symmetric matrices (GOE) and Gaussian symplectic ensemble of quaternionic Hermitian matrices (GSE) (see e.g.~\cite[\S 1.3]{Fo10} for a precise definition) exhibit structure analogous to those in the case of the GUE, and moreover in the soft edge case offer opportunities for further research.

Consider a global scaling of the eigenvalue density so that in the large $N$ limit it is supported on $(-1,1)$. The spectral moments of this limiting density again have a terminating large $N$ expansion,
\begin{equation}\label{1.1}
m_{2k}(\beta) := 
\int_{-\infty}^\infty X^{2k} 
\bar{\rho}_{(1),N}^{\rm g}(X;\beta) \, dX
= \sum_{j=0}^{2k} {a_{j}(2k;\beta) \over N^{j}}.
\end{equation}
Here, we have introduced $\beta$ --- known as the Dyson index --- as a label distinguishing the different ensembles, with $\beta = 1,2,4$ denoting the GOE, GUE, GSE respectively. In fact, this same label can be used in specifying 
$m_{2k}(\beta)$ at low order in each of these ensembles using the same formula. For example,
\begin{equation}\label{1.2}
2 m_2(\beta) = 1 + {1 \over N} \Big (-1+ {2 \over \beta} \Big ), \: \: 
2^2 m_4(\beta) = 2 + {5 \over N} \Big (-1+ {2 \over \beta} \Big ) + {1 \over N^2}\Big (3 -
{10 \over \beta} + {12 \over \beta^2} \Big );
\end{equation}
see, e.g., the recent review \cite[\S 6]{Fo25}, which discusses too the duality $m_{2k}(\beta) = m_{2k}(4/\beta) |_{N \mapsto - \beta N/2}$. We note that the duality implies that an expansion with respect to $1/N$ must be even in $N$ for $\beta = 2$, thus explaining this feature in the case of the GUE which is not present for the GOE nor GSE. Nonetheless, there are obvious structural points in common.

Said commonalities reveal themselves upon consideration of soft edge scaling of the GOE and GSE densities. However, as noticed in \cite{JM12} at the order of the first correction for the GOE, and in \cite{Bo24,Bo25a} at higher orders for the GOE and GSE, this only comes about after using in place of $N$ in the definition of soft edge scaling the shifted variable $N' := N + (\beta - 2)/(2 \beta)$. Thus, we are prescribed to define the soft edge density as
\begin{equation}\label{1.3}
\rho_{(1),N}^{s} (y;\beta)
:= {1 \over \sqrt{2} (N')^{1/6}}
\rho_{(1),N} (\sqrt{2N'} + y/(\sqrt{2} (N')^{1/6}); \beta );
\end{equation}
note that this definition is consistent with the GUE $(\beta = 2$) case as given in
(\ref{0.3}). For this rescaled quantity, it is proved in \cite{Bo25a} that for the GOE and GSE, the large $N$ asymptotic expansion is in powers of $(N')^{-2/3}\asymp N^{-2/3}$ as for the GUE (technically this is established up to the same order that the structured expansion of the GUE correlation kernel noted in the final paragraph of \S \ref{S1} is known). In contrast, the scaling \eqref{0.3} is known to result in a correction term of order $N^{-1/3}$ for the GOE and GSE. It turns out that for every choice of soft edge scaling parameters, the large $N$ expansion of the scaled density has error term of order at least $N^{-2/3}$, so \eqref{1.3} is \textit{optimal} in the sense that $\rho_{(1),N}^{s} (y;\beta)$ converges to its limit at the fastest possible rate. In further analogy with the GUE case, it is found that successive terms in this expansion are expressible in terms of (polynomial) linear combinations of  transcendental bases, now of dimension five rather than three (recall the text below \eqref{0.6}).

\section{A Research Programme}
Let us define the Gaussian $\beta$ ensemble for general $\beta > 0$ by the eigenvalue PDF proportional to 
\begin{equation}\label{2.1}
\prod_{l=1}^N e^{-\beta x_l^2}
\prod_{1 \le j < k \le N } | x_k - x_j|^\beta,
\end{equation}
where each $x_l$ (an eigenvalue) is real. This is indeed consistent with the values of $\beta$ labelling the GOE, GUE, GSE according to the Dyson index;
see, e.g., \cite[Prop.~1.3.4]{Fo10}.
In an earlier work by two of the present authors \cite{RF21}, a systematic way to derive linear differential equations of order $\beta + 1$ for the density $\rho_{(1),N}(x;\beta)$ for any $\beta$ even was given. 
By an analogue of the duality mentioned below (\ref{1.2}), after a mapping of $N$, this differential equation can equally as well be used to characterise the density for the value $4/\beta$ with $\beta$ even.
Our idea is to base a study on the asymptotic expansion of soft edge scaled densities for these $\beta$ on the differential equations. For small even $\beta$, and the corresponding value of $4/\beta$, we have found that indeed progress can be made by following this line, which to leading order (where the use of $N'$ instead of $N$ does not matter), was suggested and implemented to establish the power of $N$ for $\beta = 1,2$, and $4$ in 
 \cite[Remark 4.2]{RF21}.
We will illustrate this for $\beta = 2$.

Our starting point is the third order differential equation for the global scaled density $\rho_{(1),N}^{\rm GUE,g}(x)$,  so that the eigenvalues have leading order support $(-1,1)$,
\begin{equation}\label{2.2} 
\bigg ( \Big ( {1 \over 4 N} \Big )^2 { d^3 \over d x^3} - (x^2 - 1) {d \over dx} + x \bigg ) \rho_{(1),N}^{\rm GUE,g}(x) = 0.
\end{equation}
A comprehensive list of references of works containing a derivation of this result is given in the introduction of \cite{RF21}. These include \cite{GT05,WF14} as well as the reference \cite{RF21} --- our (\ref{2.2}) follows by setting $g=1/4, \kappa = 1$ in the first differential operator of Eq.~(2.1) of the latter reference. 
The fact that the dependence on $N$ only enters through the factor $1/(4N)^2$ is in keeping with the expansion
(\ref{0.4}) being in powers of $N^{-2}$.
Now changing variables in (\ref{2.2}) according to the soft edge scaling $x = 1 + y/2N^{2/3}$, as applies when beginning with GUE,g, we have
\begin{equation}\label{2.3} 
\Big (  { d^3 \over d y^3} - 4 y {d \over dy} + 2 \Big ) \rho_{(1),N}^{{\rm GUE,s}}(y) = {1 \over N^{2/3}} \Big (
y^2 {d \over d y} - y \Big ) \rho_{(1),N}^{{\rm GUE,s}}(y).
\end{equation}
Due to the dependence on $N$ in the coefficients of this linear differential equation only appearing via the factor
$1/N^{2/3}$, it follows immediately 
that the large $N$ asymptotic expansion of
$\rho_{(1),N}^{{\rm GUE,s}}(y)$ is in powers of this factor.

More can be said. Expanding $\rho_{(1),N}^{{\rm GUE,s}}(y) = r_0(y) + N^{-2/3}r_1(y) + N^{-4/3} r_2(y) + \cdots$ we see that successive $r_j(y)$ are related by the inhomogeneous third order differential-difference equation
\begin{equation}\label{2.4} 
\Big (  { d^3 \over d y^3} - 4 y {d \over dy} + 2 \Big )   r_{j}(y) = \Big (
y^2 {d \over d y} - y \Big ) r_{j-1}(y), \quad j=0,1,\dots 
\end{equation}
with $r_{-1}(y):=0 $ (and thus in particular, $r_{0}(y)$ satisfies the differential equation obtained by setting the RHS equal to 0 --- for references relating to this see \cite[\S 4]{RF21}).
The three linearly independent solutions of the third order differential equation given by the homogeneous part of (\ref{2.4}) can be specified. As a preliminary, one notes that the closely related third order equation $f'''(y) - 4 y f'(y) -2 f(y) = 0$ is known to have for its three linearly independent solutions $({\rm Ai}(y))^2, \, ({\rm Bi}(y))^2, \,
{\rm Ai}(y) {\rm Bi}(y)$ \cite[Eq.~(10.4.57)]{AS64}. In fact, as observed in, e.g., \cite{Na99}, setting $f = Y^2$ shows that $Y$ satisfies the Airy equation $Y''-y Y = 0$, thus explaining two of these three linearly independent solutions. In analogy, one can check that the three linearly independent solutions of the homogeneous part of (\ref{2.4}) are
 \begin{equation}\label{2.5}
 ({\rm Ai}'(y))^2 - y ({\rm Ai}(y))^2, \quad
  ({\rm Bi}'(y))^2 - y ({\rm Bi}(y))^2, \quad
 {\rm Ai}'(y) {\rm Bi}'(y) - y {\rm Ai}(y)   {\rm Bi}(y),
 \end{equation}
 with the first equalling $r_{0}(y)$; recall (\ref{0.3}).
In relation to (\ref{2.4}) for general $j$, it is simple to check (inductively) that the particular solutions are of the form noted in the
 final paragraph of \S \ref{S1}. 

 A simplification of \eqref{2.4} is possible. Thus,
 multiply both sides by $e^{\gamma y}$, integrate over $y \in (-\infty,\infty)$
 (this requires $\gamma>0$ for convergence), and simplify using integration by parts to deduce that the Laplace transforms $u_j(\gamma):=\int_{-\infty}^{\infty}e^{\gamma y}r_j(y)\, dy$ satisfy the first order inhomogeneous differential-difference equation
\begin{equation}\label{2.6}
4\gamma u_j'(\gamma)+(6-\gamma^3)u_j(\gamma)=-\gamma u_{j-1}''(\gamma)-3u_{j-1}'(\gamma), \quad u_{-1}:=0.
\end{equation}
It is known in the literature
\cite{Ok02,OS20}, \cite[Eq.~(3.54)]{Fo21} that the exact functional form of the Laplace transform of the GUE soft edge scaled density is
$
u_0(\gamma)=e^{\gamma^3/12}/(2\sqrt{\pi}\gamma^{3/2})$, which we can check solves the homogeneous part of (\ref{2.6}). The next line of study on this is to use (\ref{2.6}) inductively to characterise each of the $u_j(\gamma)$ for $j>0$.

 Going forward from here, one foresees a few lines of research to pursue: The simplest of these is to extend the relevant calculations to other low values of even $\beta$, and the corresponding $4/\beta$,
starting with $\beta = 4$ where from \cite{RF21} we have an explicit fifth order differential equation available.
Rather than considering cases, one must also wonder if the theory of \cite{RF21} can be applied to general even $\beta$, where we have available a matrix differential equation characterisation, but not the explicit scalar differential equation.
For general even $\beta$ there is also a $\beta$-dimensional integral formulation, which has been used in earlier studies to compute the limiting and first order correction
\cite{DF06,FT19a}.
Finally, the methods described in this note also apply to the Laguerre and Jacobi ensembles of equivalent $\beta$
from the theory of \cite{RF21}. The limiting soft edge scaled densities of said ensembles are known to agree with those in the Gaussian case. It remains to see how replacing $N$ by $N'=N+(\beta-2)/(2\beta)$ in the traditional soft edge scaling variables for these ensembles affects the asymptotic expansions at the soft edge. We have available \cite{Bo25a} for guidance in the Laguerre cases with $\beta=1,2$, and $4$.

\section*{Acknowledgement}
The work of Peter J. Forrester is supported by the Australian Research Council Discovery Project
DP250102552.
 Bo-Jian Shen is supported by the Australian Research Council Discovery Project DP210102887 and the Shanghai Jiao Tong
University Overseas Joint Postdoctoral Fellowship Program. Anas A. Rahman is supported by Hong Kong GRF 16304724 and NSFC 12222121 and benefited from a visit to the University of Melbourne funded by the MATRIX--Simons Young Scholar Program.
The authors acknowledge helpful correspondence on our proposed research programme from Folkmar Bornemann.
The authors thank the organisers of the MATRIX program ``Log-gases in Caeli Australi'', held in Creswick, Victoria, Australia during the first half of August 2025 
for their efforts in bringing together many national and international researchers in the field, and for
facilitating our collaboration.

  \end{document}